# Ultrabright narrow-band telecom two-photon source for long-distance quantum communication

Kazuya Niizeki[1], Kohei Ikeda[1], Mingyang Zheng[2], Xiuping Xie[2], Kotaro Okamura[3], Nobuyuki Takei[4,5], Naoto Namekata[6], Shuichiro Inoue[6], Hideo Kosaka[1], and Tomoyuki Horikiri[1,7,*]

[1]*Yokohama National University, Yokohama 240-8501, Japan*

[2]*Jinan Institute of Quantum Technology, Jinan, Shandong 250101, China*

[3]*Kanagawa University, Yokohama 221-8686, Japan*

[4]*Institute for Molecular Science, Okazaki, Aichi 444-8585, Japan*

[5]*SOKENDAI (The Graduate University for Advanced Studies), Okazaki, Aichi 444-8585, Japan*

[6]*Nihon University, Chiyoda, Tokyo 101-8308, Japan*

[7]*JST, PRESTO, Kawaguchi, Saitama 332-0012, Japan*

*E-mail: horikiri-tomoyuki-bh@ynu.ac.jp

We demonstrate an ultrabright narrow-band two-photon source at the 1.5 -μm telecom wavelength for long-distance quantum communication. By utilizing a bow-tie cavity, we obtain a cavity enhancement factor of $4.06 \times 10^4$. Our measurement of the second-order correlation function $G^{(2)}(\tau)$ reveals that the linewidth of 2.4 MHz has been hitherto unachieved in the 1.5 -μm telecom band. This two-photon source is useful for obtaining a high absorption probability close to unity by quantum memories set inside quantum repeater nodes. Furthermore, to the best of our knowledge, the observed spectral brightness of $3.94 \times 10^5$ pairs/(s·MHz·mW) is also the highest reported over all wavelengths.

Quantum communication is attractive for its wide range of applications, for example, it enables unconditionally secure communication via quantum key distribution (QKD), realizes a precise "global" clock by connecting many atomic/optical clocks existing at various nodes all over the world,[1] and enables cloud and/or distributed quantum computing.[2] Currently, the longest quantum communication distance achieved via the use of optical fiber channels is around 400 km.[3] To further increase the distance, quantum repeaters enabling long-distance entanglement distribution are necessary.[4-6] The following are needed to implement quantum repeaters: a quantum light source with emission at telecom wavelengths to ensure low-loss propagation in optical fibers, quantum memory to preserve photonic quantum states in static matter qubits, and wavelength conversion between telecom wavelength and quantum memory wavelength.[7,8]

For the light source, two-photon sources including polarization, time-bin, and/or frequency entanglement have been implemented by spontaneous parametric down-conversion (SPDC). SPDC used as an entangled photon-pair source as well as a heralded single-photon source[9,10] typically has a large spectral width of >10 nm because of the nonlinear phase-matching condition. On the other hand, most quantum memories have narrow linewidths, ranging from kilohertz order for the homogeneous linewidths of rare-earth-ion-doped solids (REIDS) to gigahertz order for the inhomogeneous linewidths of quantum dots and atomic gas. Thus far, narrow-band two-photon sources using cavities have been developed to achieve enhanced two-photon generation rates at certain wavelengths, including near-infrared[11-16] and telecom bands.[17-21] However, in most cases, wavelengths of the generated photons must be converted to match the quantum memory wavelengths. Wavelength conversion techniques[7,8,22-26] have been developed both for high-efficiency single-photon detection of telecom photons[22] and for connecting telecom wavelengths and the wavelengths of quantum memories, including InAs quantum dots (900 nm)[7] and NV centers in diamond (637 nm).[26]

For atomic frequency comb (AFC) memories,[5,27] which require tailored absorption peaks of REIDS for collective emission, a narrow linewidth photon source can be used for wavelength multiplexing because of the increased number of narrow range channnels (note that one range corresponds to one wavelength channel) in an inhomogeneous broadened spectrum. Although, the inhomogeneous linewidth is nearly 10 GHz[28] for the $Pr^{3+}$:YSO crystal for instance, AFC memory has some spin levels that limit each available range to nearly 10 MHz. This means that the linewidth of the photon source must be < 10 MHz even without wavelength multiplexing. The recall efficiency of AFC memory can theoretically be made nearly unity for small linewidth photon by optimizing the finesse of the absorption peaks.[27]

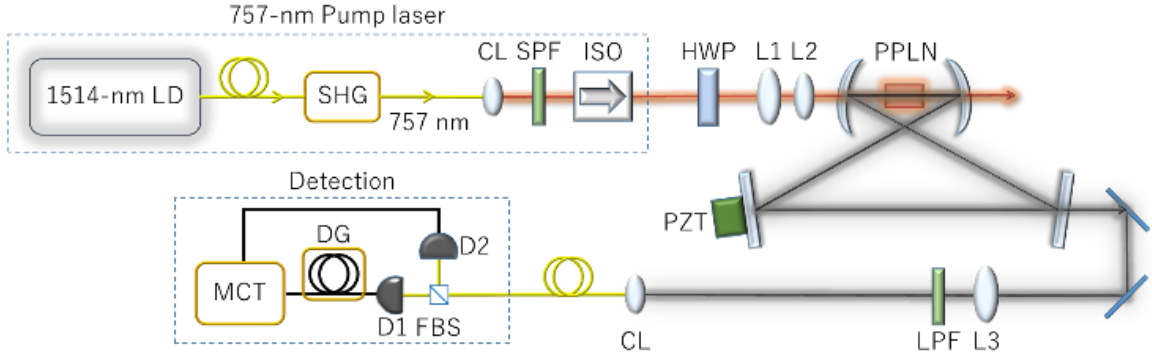

Fig. 1. Experimental setup of two-photon source system. LD, laser diode; SHG, fiber SHG module using PPLN; CL, collimate lens; SPF, short-pass filter; ISO, isolator; HWP, half-wave plate; L1, lens with $f = 100$ mm; L2, lens with $f = 50$ mm; L3, lens with $f = 400$ mm; PZT, piezoelectric transducer; LPF, long-pass filter; FBS, fiber beam splitter; D1 and D2, avalanche photodiodes; DG, delay generator; MCT, multichannel timer.

In this Letter, we report the development of an ultrabright, narrow-band telecom wavelength two-photon source that provides a coupling efficiency of nearly unity with quantum memories when combined with the above-mentioned wavelength conversion technique. By utilizing a cavity-enhanced SPDC, we demonstrate, to the best of our knowledge, the narrowest linewidth for the 1.5 -μm telecom wavelength and the highest ever spectral brightness.

Our experimental setup is shown in Fig. 1. The master laser in the setup is an amplified external cavity diode laser (Sacher TEC420-1530-1000) that can be scanned over the telecom wavelength range. We select 1514 nm for demonstration in this article, because the sum frequency generation of 1514 nm (the present photon source) and 1010 nm (the nominal pump laser), which are respectively stabilized by the saturated absorption spectroscopy of acetylene and iodine, is 606 nm and matches the $^3H_4 - {}^1D_2$ transition of the $Pr^{3+}$:YSO crystal being utilized for memory transition.

The 1514-nm output is transmitted to a periodically poled lithium niobate (PPLN) waveguide for second-harmonic generation (SHG). This 757-nm SHG light is used as the pump laser for SPDC. The output from the PPLN waveguide is collimated and transmitted through short-pass filters (Thorlabs, FESH0800) to eliminate the remaining telecom wavelength light and the half-wave plate in order to optimize polarization. Then, the 757-nm beam passes through a lens pair with focal lengths of 100 and 50 mm and is then focused onto a PPLN crystal chip for SPDC (type-0, 10-mm length), located at the focused beam waist position inside the cavity. A degenerate 1.5-µm photon pair is generated by temperature phase matching. The chip temperature is controlled to ~300 K with an accuracy of 0.01 K. The beam waist size in the PPLN chip in the short arm of the cavity (Fig.1) is around 20 µm to optimize parametric conversion efficiency.[29] The PPLN surfaces are antireflection-coated for telecom wavelengths to minimize loss inside the cavity. The cavity is a bow-tie cavity with a round trip length of around 60 cm. The two mirrors in the short arm are concave mirrors (radius of curvature = 50 mm), while the mirrors in the long arm are flat mirrors. At telecom wavelengths, the concave mirrors and one of the flat mirrors have a reflectivity $R$ of > 99.9%, while the other flat mirror is used as an output coupler and has $R$ = 95 or 99% for two different cavity-finesse setups. The pump wavelength reflectivity is < 2% for all the mirrors. A piezoelectric element is attached to the back of one flat mirror to scan the cavity resonant frequency and/or to lock the cavity length by the Pound-Drever-Hall technique. In this study, we apply an alignment beam (1.5-µm wavelength) for the cavity and use its reflection to estimate loss inside the cavity; the cavity loss except for the output mirror is estimated to be ~1.7%. Subsequently, the SPDC linewidths for $R$ = 95 or 99% are estimated as 5.6 and 2.2 MHz, respectively. After traveling through long-pass filters (Thorlabs FELH1400) to remove the pump photons, the SPDC photons from the cavity enter a Hanbury–Brown–Twiss-type setup for the measurement of the second-order cross-correlation function $G^{(2)}(\tau)$. The photon detectors (Princeton Lightwave, PGA-016u-1550TFX) used in the measurement are fiber-coupled-type avalanche photodiodes (APDs) working in the passive quench mode. Their dark count rates are around 6.5 kHz, and the detection efficiency is 5%. APDs are operated in the Geiger mode using conventional passive quenching circuits. Although the Geiger APD realizes a free-running (nongated) single-photon detection, it produces noise counts (dark counts and afterpulses) higher than those in the gate operation.[30] In this work, the detection efficiency is restricted by 5% to reduce the noise counts (the efficiency and noise counts have a strong trade-off relationship). The detector output signals are sent to a multi-channel timer (PicoQuant, Hydraharp 400).

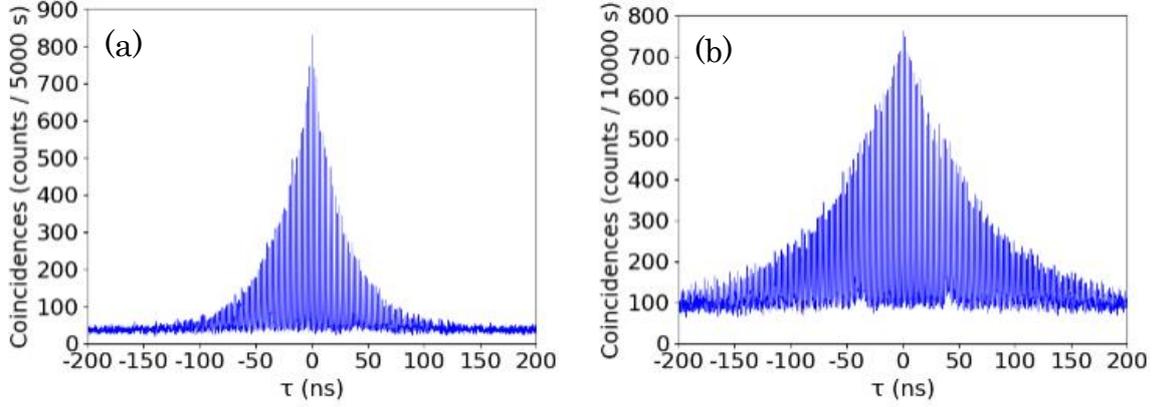

Fig. 2. Two-photon coincidence counts for different reflectivities of cavity output mirror. (a) $R$ = 95% at 5-μW pump power with 4.5-kHz single count rate for 5000 s and (b) $R$ = 99% at 10-μW pump power with 4.7-kHz single count rate for 10000 s. Measurement parameters: 128-ps time bin size, 5% detector efficiency with 6.5-kHz dark count.

In the following, we show the linewidth and spectral brightness of the present two-photon source. For cavity-enhanced SPDC photons that are pumped far below the oscillating threshold,[31] the measured $G^{(2)}(\tau)$ provides information of the linewidth, free spectral range, and mode numbers.[19] The correlation function decays with increasing round-trip numbers owing to cavity loss and the linewidth for the generated two-photon pair can therefore be deduced. The spectral brightness is estimated from the linewidth, mode number, and pump power, as shown later.

In $G^{(2)}(\tau)$, an additional comb structure is created because of the multimode nature of the two-photon source.[12] Figure 2 shows the measured $G^{(2)}(\tau)$, wherein Fig. 2(a) corresponds to the case of $R$ = 95% and pump power of 5 μW and Fig. 2(b) corresponds to $R$ = 99% and pump power of 10 μW. The measurement times are 5000 and 10000 s, respectively. Accidental coincidence counts due to dark count and/or stray light are not subtracted. The period of the comb structure corresponds to the cavity round-trip time, and the decay rate of the envelope corresponds to the cavity linewidth. A higher reflectivity [Fig. 2(b)] yields a narrower linewidth and a lower signal-to-noise ratio. This is because the average round trip number is higher, and the resulting escape efficiency for $R = 99\%$ decreases owing to higher internal loss.

The parameter $G^{(2)}(\tau)$ from the cavity two-photon source can be expressed as[12]

$$G^{(2)}(\tau) = \langle E^-(\tau)E^-(t+\tau)E^+(t+\tau)E^+(\tau)\rangle$$
$$= C_1\left[e^{-\omega_W \tau}\left|\frac{\sin[(2N+1)\Delta\Omega_F \tau/2]}{\sin(\Delta\Omega_F \tau/2)}\right|^2 + C_2\right] \quad (1)$$

Here, $E^-(\tau)$ and $E^+(\tau)$ denote the creation and annihilation operators of a down-converted photon, respectively, $\omega_W$ is the cavity linewidth (FWHM), $\Delta\Omega_F$ is the free spectral range (FSR), N is the longitudinal mode number, and the constants $C_1$ and $C_2$ depend on pump power.

From Eq. (1), we note that the decay of the envelope yields the cavity linewidth. In actual measurement, the timing jitter $J(\tau')$ of the detection system also needs to be considered. Via the convolution of the jitter for two detectors, the detected second-order correlation function $G_c^{(2)}(\tau)$ can be expressed as

$$G_c^{(2)}(\tau) = C_1[e^{-\omega_W \tau}\Sigma_n \int d\tau' J(\tau')J(\tau' + \tau - n\tau_F) + C_2]$$
$$= C_1\left[e^{-\omega_W \tau}\sum_{n=0}^{\infty}\exp\left(-\frac{4\ln 2\,(\tau - n\tau_F)^2}{\tau_W^2}\right) + C_2\right] \quad (2)$$

Here, $\tau_F = 2\pi/\Delta\Omega_F$ represents the cavity round trip time and $\tau_W$ is the temporal width (FWHM) of one tooth of the comb. In the calculation, the ratio of the two sine functions in Eq. (1) is replaced with delta functions $\Sigma_n \delta(\tau - n\tau_F)$, and jitter is assumed to be the Gaussian function $\exp\left(-\frac{8\ln 2\,\tau^2}{\tau_W^2}\right)$. The jitter of a single detector is given by $\tau_W/\sqrt{2}$, and this value is typically 300 ps.

The fitting results obtained with Eq. (2) are shown in Fig. 3. The deviation between the data and the fitting is due to noise fluctuation. The obtained linewidths are (a) 5.3 and (b) 2.4 MHz. Furthermore, the separation between the teeth is $\tau_F$ =1.9 ns, corresponding to a cavity length of 57 cm and FSR of 526 MHz. The finesse values are (a) 99 and (b) 220. These values are within an error of 10% from the calculation results of the cavity design.

The comb structure in Fig. 2 means that the two-photon spectrum obtained from the cavity consists of multiple longitudinal modes, since the comb structure will disappear if there is only one single longitudinal mode.[19] Furthermore, the ratio of $\tau_F$ to $\tau_W$ is equal to the number of longitudinal modes involved. However, the measured $\tau_W$ also includes the timing jitter of the detection system. Therefore, timing jitter must be determined to estimate the longitudinal mode numbers. To determine the timing jitter of our measurement system, we measure the coincidence counts of SPDC photon pairs with the output mirror removed (single-pass measurement), wherein the temporal width of the coincidence peak yields the timing jitter of the system. The obtained value is 509 ps, and under the assumption that the temporal distribution is Gaussian, the jitter of one detector is calculated as ~350 ps. The fitting results in Fig. 3 yield (a) $\tau_W = 528$ ps and (b) $\tau_W = 561$ ps. Because of the additivity of dispersion, temporal widths corresponding to multiple longitudinal modes are (a) 140 ps, (b) 236 ps. The ratio of $\tau_F$ to $\tau_W$ corresponds to 2N+1, where N denotes the mode number,[13,31] and therefore the mode numbers in our study are (a) ~ 6 and (b) ~ 3. Precise evaluation of the mode numbers is possible with the introduction of an additional scanning cavity, as demonstrated in Ref. 21.

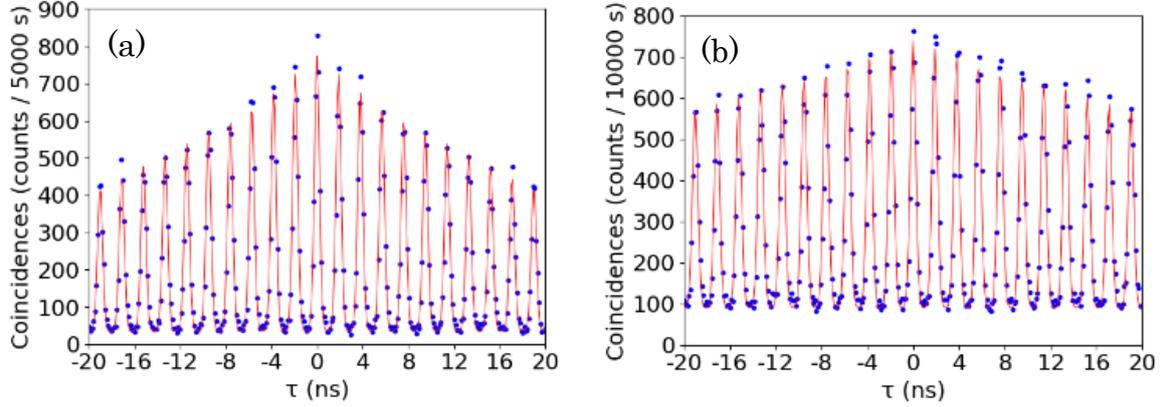

Fig. 3. Fitting results of the data around the center (delay zero) in Fig. 2. The blue dots correspond to the ones in Fig. 2, and the red solid lines denote the fitting curve of Eq. (2). Fitting parameters: (a) $C_1 = 738$, $C_2 = 0.048$, $\tau_F = 1.9$ ns, $\tau_W = 528$ ps, $\omega_W/2\pi = 5.3$ MHz and (b) $C_1 = 650$, $C_2 = 0.14$, $\tau_F = 1.9$ ns, $\tau_W = 561$ ps, $\omega_W/2\pi = 2.4$ MHz.

$g^{(2)}(0)$ is another important parameter when the present two-photon source is connected to quantum memory. It is the normalized cross-correlation at zero delay.[19] Figure. 4 depicts the pump-power dependence of $g^{(2)}(0)$. We note that the data shown were obtained separately from the data in Fig. 2. In the relatively strong pump power regime, $g^{(2)}(0)$ decreases while the background noise increases. This background is mainly attributed to simultaneous multiple photon-pair generations in the cavity, which becomes noise when this source is utilized in quantum communication. At lower pump powers of ~5-10 μW, $g^{(2)}(0)$ exhibits maximum values and decreases with the decrease in the pump power owing to the dark counts of the single-photon detectors.

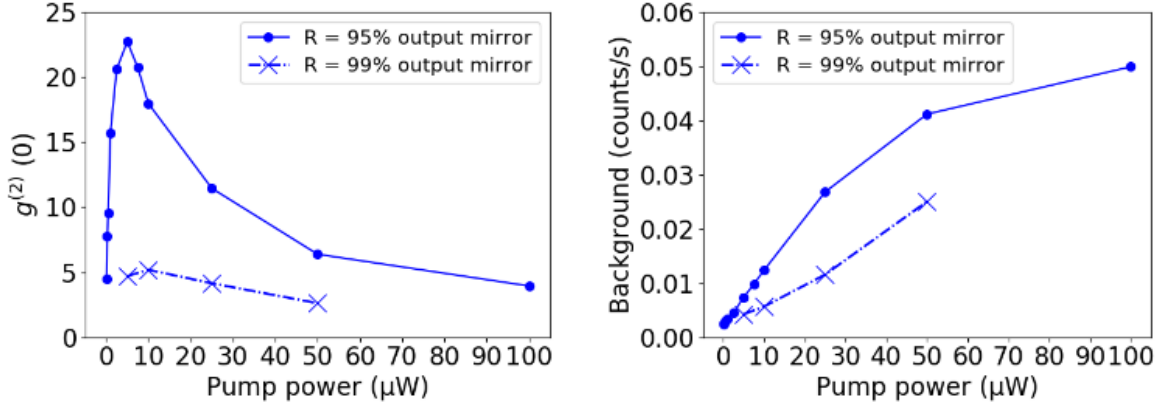

Fig. 4. (left) Parameter $g^{(2)}(0)$ and (right) background floor as functions of pump power (minimum 0.1 μW). The solid line linking the dots corresponds to $R = 95\%$ of the output mirror for 1000 s per measurement, and broken line linking the crosses corresponds to $R = 99\%$ for 5000 s per measurement. Other measurement parameters are the same as in Fig. 2.

To estimate the spectral brightness, we calculate the coincidence counts by summing $G^{(2)}(\tau)$ without background noise.[16,20] From Fig. 2, we note that the total coincidence counts are (a) $1.05 \times 10^5$ and (b) $2.07 \times 10^5$ pairs/10000 s. Dividing these by the mode number, linewidth, and pump power, $R_{detect}$ = (a) 132 and (b) 288 pairs/(s·MHz·mW) are obtained. These values include losses in the present measurement. The spectral brightness inside the cavity $R_{generation}$ is calculated from

$$R_{generation} = R_{detect}/(t_1 f t_2 d)^2 \quad (3)$$

and the parameters describing losses are summarized in Table 1.

The spectral brightness values obtained using the values in Table 1 are (a) $1.81 \times 10^5$ and (b) $3.94 \times 10^5$ pairs/(s·MHz·mW). These values are, to the best of our knowledge, the highest reported thus far. Table 2 lists the observed trends for cavity-enhanced narrow-band two-photon sources.

Table 1. Efficiency of detection system

| Symbol | Meaning | Efficiency |
|:---:|:---:|:---:|
| $t_1$ | Transmittance of optical element to SMF | 96% |
| $f$ | Coupling efficiency of SMF | 58% |
| $t_2$ | Transmittance after FBS | 97% |
| $d$ | APD efficiency | 5% |

Table 2. Recent trends of cavity photon sources

| Reference | Wavelength (nm) | Bandwidth (MHz) | Spectral brightness $(s \cdot MHz \cdot mW)^{-1}$ |
|:---:|:---:|:---:|:---:|
| Fekete[19] | 1436 (606) | 1.7 (2.9) | $8.00 \times 10^3$ |
| Zhou[20] | 1560 | 8.0 | $1.34 \times 10^2$ |
| Rambach[15] | 795 | 0.66 | $3.58 \times 10^4$ |
| Tsai[16] | 852 (780) | 6.0 (6.6) | $1.06 \times 10^5$ |
| This work | 1514 | 2.4 | $3.94 \times 10^5$ |

A comparison of the above results with the data of the single-pass SPDC measurement, obtained by removing the output coupler, reveals that cavity enhancement factors are (a) $1.86 \times 10^4$ and (b) $4.06 \times 10^4$, which are of the same order of the square of finesse[31]: (a) $9.80 \times 10^3$ and (b) $4.84 \times 10^4$.

Our high spectral brightness and enhancement factor are explained as follows. Firstly, we use type-0 quasi-phase-matching PPLN, which has the highest nonlinear coefficient. For type-0, -1, and -2 SPDC in the articles cited in Table 2, the related nonlinear coefficients are $d_{33}, d_{31}, and\ d_{24}$, respectively, and generally. $d_{33}$ is about one order of magnitude higher than the others (for instance, $d_{33,PPLN} \sim 30, d_{31,PPLN} \sim 6,\ and\ d_{24,PPKTP} \sim 3$, in units of pm/V; these values vary slightly with wavelength). Secondly, frequency-degenerate SPDC has a slightly higher generation rate than non-degenerate SPDC because both beams overlap better. Thirdly, the loss inside the telecom cavity is lower than when shorter wavelengths are generated since telecom wavelength does not cause SPDC loss such as green-light-induced infrared absorption. Lastly, narrow linewidth has a great influence on spectral brightness by definition, showing its merit for quantum communication. Of course, in the future, the spectral brightness can be further increased by optimizing the system parameters.

In conclusion, we demonstrated a cavity-enhanced telecom narrow-band two-photon source with the highest spectral brightness and narrowest linewidth around of 1.5 μm. This source is a promising candidate for long-distance quantum communication where efficient coupling with quantum memory is necessary. Furthermore, the availability of the frequency and time-bin entanglement of this source[32] can also be utilized in quantum information processing while polarization entanglement can be easily implemented by utilizing two PPLN chips[14].

**Acknowledgments**

This work was supported by the Toray Science foundation, the Asahi Glass foundation, KDDI foundation, the Murata Science foundation, JKA, REFEC, SECOM foundation, JST PRESTO JPMJPR1769, and JST START ST292008BN, Japan.
We thank H. Goto for his helpful comments and Q. Zhang, Y. Yamamoto, S. Utsunomiya, T. Kobayashi, M. Fraser, and I. Iwakura for their support.